\documentclass[12pt,preprint]{aastex}
\usepackage{natbib}

\shorttitle{Astrometry and RV of GJ 317}

\shortauthors{Anglada-Escud\'e, Boss, Weinberger, Butler, et al.}

\begin{document}

\title{Astrometry and radial velocities of the planet host M dwarf GJ 317: new
trigonometric distance, metallicity and upper limit to the mass of GJ 317b}

\author{
Guillem Anglada-Escud\'e\altaffilmark{1}, 
Alan P. Boss\altaffilmark{1}, 
Alycia J. Weinberger\altaffilmark{1},
Ian B. Thompson\altaffilmark{2},
R. Paul Butler\altaffilmark{1},
Steven S. Vogt\altaffilmark{3},
Eugenio J. Rivera\altaffilmark{3},
}
\email{
anglada@dtm.ciw.edu, 
}

\altaffiltext{1}{Department of Terrestrial Magnetism, Carnegie Institution for
Science, 5241 Broad Branch Road NW, Washington D.C., 20015, USA}
\altaffiltext{2}{Carnegie Observatories, 813 Santa Barbara Street, Pasadena, CA
91101, USA}
\altaffiltext{3}{UCO/Lick Observatory, University of California, Santa Cruz, CA
95064, USA}

\begin{abstract} 

We have obtained precision astrometry of the planet hosting M dwarf GJ
317 in the framework of the Carnegie Astrometric Planet Search project.
The new astrometric measurements give a distance determination of 15.3
pc, 65\% further than previous estimates. The resulting
absolute magnitudes suggest it is metal rich and more massive than
previously assumed.  This result strengthens the correlation between
high metallicity and the presence of gas giants around low mass stars.
At 15.3 pc, the minimal astrometric amplitude for planet candidate GJ
317b is 0.3 milliarcseconds (edge-on orbit), just below our astrometric
sensitivity. However, given the relatively large number of observations
and good astrometric precision, a Bayesian Monte Carlo Markov Chain analysis
indicates that the mass of planet b has to be smaller than twice the
minimum mass with a 99\% confidence level, with a most likely value of
2.5 $M_{Jup}$. Additional RV measurements obtained with Keck by the
Lick-Carnegie Planet search program confirm the
presence of an additional very long period planet candidate, with a
period of 20 years or more. Even though such an object will imprint a
large astrometric wobble on the star, its curvature is yet not evident
in the astrometry. Given high metallicity, and the trend indicating
that multiple systems are rich in low mass companions, this system is
likely to host additional low mass planets in its habitable zone that
can be readily detected with state-of-the-art optical and near infrared
RV measurements. 
\end{abstract}


\section{Introduction}

Astrometric observations complementing radial velocity
measurements provide the only means to measure the dynamical
masses of long period non-transiting exoplanets. The
astrometric technique has been largely unsuccessful in the
detection of low mass companions, mainly due to strong
systematic effects that are difficult to calibrate through the
noise added by the Earth's atmosphere. An example is the planet
candidate around the M8.5V star VB 10 \citep{pravdo:2009},
which could not be confirmed by radial velocities
\citep{bean:2010,anglada:2010} nor by further astrometric
observations \citep{lazorenko:2011}. Compared to other
techniques such as precision Doppler measurements, astrometry
cannot be corrected using laboratory standards and relies upon
measuring the relative motion of the target star with respect
to a number of background and (desirably) slow moving distant
sources. Typical FGK planet hosting dwarfs that should show the
largest astrometric wobbles are too bright to be observed
simultaneously with  background sources, so the perturbing
effect of the atmosphere and time dependent geometric field
distortions cannot be easily calibrated in the observations.
Some success has been achieved from space. Posterior analysis
of HIPPARCOS astrometry has been used to put upper limits on
the masses of detected RV companions, ruling out that most of
them are low mass stars in face on orbits
\citep{pourbaix:2001}. A recent reanalysis by
\citet{reffert:2011} of the residuals to the new HIPPARCOS
solution \citep{hipparcos:2010} has been able to confirm the
planetary nature of 9 objects and a number of brown dwarf
candidates by putting upper limits to their masses. In the same
study, the planetary signal could be recovered on 3 systems.
Also, precision astrometry of bright stars has been measured
using the Fine Guidance Sensors on board the Hubble Space
Telescope \citep[e.g.][ for the most recent
results]{martioli:2010,mcarthur:2010}. Up to now, these are the
only astrometric measurements sensitive enough to place actual
constraints on the masses of long period planet candidates and
are all obtained by space-based observations.

The Carnegie Astrometric Planet Search is a ground-based program
focused on the detection of giant planets around nearby, low mass
stars. The astrometric wobble imprinted by a planet on the central
star is inversely proportional to the stellar mass (the smaller the
mass, the larger the signal). Also, low mass stars (M and L dwarfs)
are much fainter than typical F,G and K stars (the most common
targets of the RV surveys), simplifying the simultaneous
observations of the targets and faint background reference sources.
A central part of the project has been the construction of a
specialized camera \citep{boss:2009}. CAPSCam uses a Hawaii-2RG
HyViSI hybrid array that allows the definition of an arbitrary
guide window which can be read out (and reset) rapidly, repeatedly,
and independently of the rest of the array. This guide window is
centered on our relatively bright target stars, with multiple short
exposures avoiding saturation. The rest of the array then
integrates for prolonged periods on the background reference grid
of fainter stars. This dramatically extends the dynamic range of
the composite image. This HyViSI detector is the heart of the
CAPSCam concept. A full description of the program, instrument
performance and characteristics is given in \citet{boss:2009}. The
high dynamical range enabled by the guide window mode permits the
observation of stars as bright as magnitude I~=~9 using integration
times as short at 0.2 sec on the target object. Even though the
best epoch precision is at the level of 0.4 mas, the long term
accuracy is limited by (yet) unknown systematic effects, that are
more pronounced on bright objects. The typical epoch-to-epoch
precision at the bright end is about 1 mas. As a
complementary part of the planet search program, we follow-up a
handful of long period RV planet candidates that should produce a
detectable astrometric signal (that is, with a semiamplitude above
0.5 mas). GJ 317 is one such star  
\citep[I~$\sim$~9.32,][]{rojo:2003}. \citet{johnson:2007}
(hereinafter JB07) reported the detection of a $\sim$700 day period
planet candidate with a minimum mass of 1.2 $M_{Jup}$ and evidence
of an additional companion (GJ 317c) with a period of several
thousand days. Using that orbital solution and the previously
reported distance  \citep[$\sim$9.7 pc,][]{jenkins:1963}, we
estimated a minimum astrometric amplitude (assuming an edge-on
orbit) of $\sim$0.58 mas for GJ~317b. With an epoch-to-epoch
precision of 1 mas, this is a challenging but doable measurement.
Assuming a median inclination of 45 deg, one would expect a
semiamplitude of 0.84 mas.

The star has been continuously monitored by the Lick-Carnegie Planet
Search group using Keck/HIRES, increasing the number of RV measurements
from 18 (JB07) to 38. These additional measurements should better
constrain the nature of the long period planet and permit a search for
lower amplitude signals. GJ 317 is a nearby star of astrobiological
interest. Its low mass and Solar System-like configuration (two outer
giant planets) make it an outstanding candidate to search for potentially
habitable planets with radial velocity observations.

\section{Observations: Astrometry}
\begin{figure}[tb]
\center
\includegraphics[angle=0, width=5in, clip]{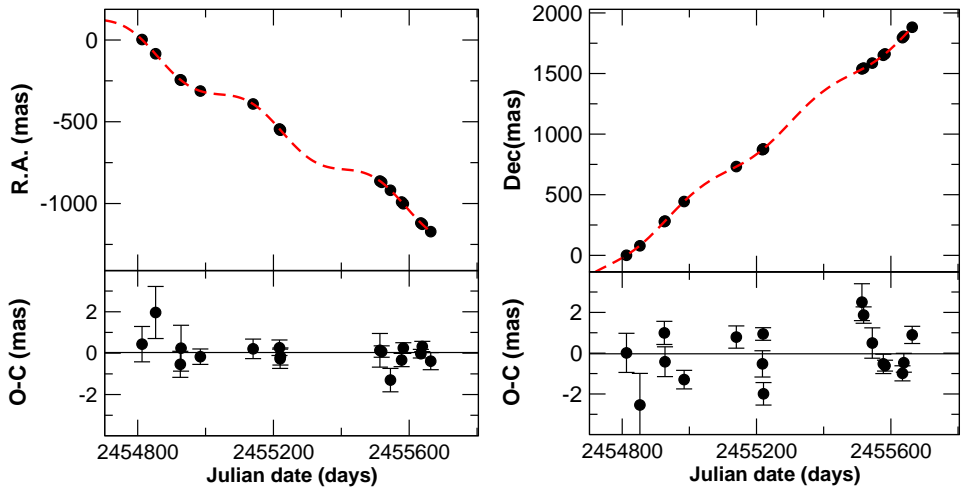}

\caption{Top panels. Differential astrometry of GJ 317 measured with
CAPScam as a function of time. The best fit to the
parallax and proper motion is illustrated by the red dashed line. Bottom
panels. The residuals after removing the parallax and 
proper motion.
}
\label{fig:rawastrometry}
\end{figure}

We have used CAPSCam mounted on the 2.5-m du Pont telescope at Las
Campanas  Observatory to obtain 18 astrometric observations   of
GJ~317 with a time baseline of 2.5 years. The observations cover 1.5
orbital periods of GJ 317b. Since the RV measurements constrain  5
out of 7 of the orbital parameters (period $P$, minimum mass $M \sin
i$, initial mean anomaly $M_0$, eccentricity $e$, and argument of the
periastron $\omega$), the astrometric measurements only need to
constrain the two remaining elements (argument of the node $\Omega$
and orbital inclination $i$), plus the classic astrometric
observables (position, proper motion $\mu_\alpha$ and $\mu_\delta$;
and the parallax $\Pi$, see Appendix \ref{sec:processing}).

A typical observing epoch consists spending one hour of telescope
time to obtain 30-50 images on the full $\sim 6\prime \times
6\prime$. The integration time of each image varies between 45 to
60 seconds depending on seeing and environmental conditions.
Guide-window exposure time is always set to 0.2 seconds on GJ
317. Therefore, between 225 and 300 guide-window reads are
obtained and coadded to each image of the full field.  The field
around GJ 317 is rich in background reference stars. The
astrometric precision (intranight centroid scatter) is computed
by comparing the position of the target star to the reference
stars in each of the 30-50 images per epoch. The epoch-to-epoch
accuracy is estimated by comparing the observations to the best
fit astrometric model (see Appendix \ref{sec:processing}). For GJ
317 in particular, the typical intranight precision is of the
order of 0.5 mas. This precision is illustrated as the error bars
in Figure \ref{fig:rawastrometry}. However, the typical
epoch-to-epoch accuracy is 0.6 mas in R.A. and 1.2 mas in
declination, indicating that a significant source of uncertainty
is not random errors but uncalibrated field distortion effects
(systematic errors), especially in Declination. One epoch
(January 2009) was rejected because the camera shutter stopped
working, causing severe saturation and charge persistence
problems on the target star. The relative astrometric
measurements used in the analysis are given in Table
\ref{tab:astrodata} and illustrated in Figure
\ref{fig:rawastrometry}. The residuals (bottom panels) are
obtained by subtracting the best fit to the parallax and proper
motion. An overview of the astrometric solution and relevant
statistical quantities are given in Section \ref{sec:analysis}.
Further details on the astrometric data reduction procedure are
given in Appendix \ref{sec:processing}.

\begin{deluxetable}{lcccc}
\tablecaption{Differential astrometry of GJ 317
\label{tab:astrodata}}
\tablehead{
\colhead{Julian date (days)} & 
\colhead{R.A.} & 
\colhead{Dec.} & 
\colhead{Unc. R.A.} & 
\colhead{Unc. Dec.}\\
\colhead{(days)} & 
\colhead{(mas)} & 
\colhead{(mas)} & 
\colhead{(mas)} & 
\colhead{(mas)}
}
\tablewidth{0pt}
\startdata
2454812.73956  &     2.53  &	 0.45	&    0.85   &	 0.96  \\
2454852.70285  &   -84.44  &	78.60	&    1.25   &	 1.54  \\
2454925.62847  &  -243.32  &   278.06	&    0.62   &	 0.56  \\
2454927.64450  &  -245.94  &   282.56	&    1.10   &	 0.73  \\
2454984.47159  &  -312.21  &   443.92	&    0.37   &	 0.45  \\
2455139.85431  &  -391.52  &   732.33	&    0.47   &	 0.54  \\
2455217.72615  &  -543.92  &   871.29	&    0.37   &	 0.64  \\
2455219.71792  &  -549.13  &   877.34	&    0.46   &	 0.30  \\
2455220.72072  &  -551.37  &   876.73	&    0.41   &	 0.55  \\
2455513.86011  &  -862.40  &  1538.05	&    0.81   &	 0.90  \\
2455518.86210  &  -870.26  &  1544.93	&    0.27   &	 0.40  \\
2455544.83951  &  -919.02  &  1586.19	&    0.56   &	 0.74  \\
2455577.73092  &  -990.53  &  1650.71	&    0.32   &	 0.47  \\
2455582.66573  & -1001.48  &  1661.73	&    0.25   &	 0.26  \\
2455634.52079  & -1118.87  &  1795.91	&    0.20   &	 0.37  \\
2455638.55749  & -1126.62  &  1807.95	&    0.24   &	 0.46  \\
2455663.52245  & -1171.86  &  1882.12	&    0.41   &	 0.42  \\ 
\enddata
\end{deluxetable}

\begin{figure}[tb]
\center
\includegraphics[angle=0, width=5in, clip]{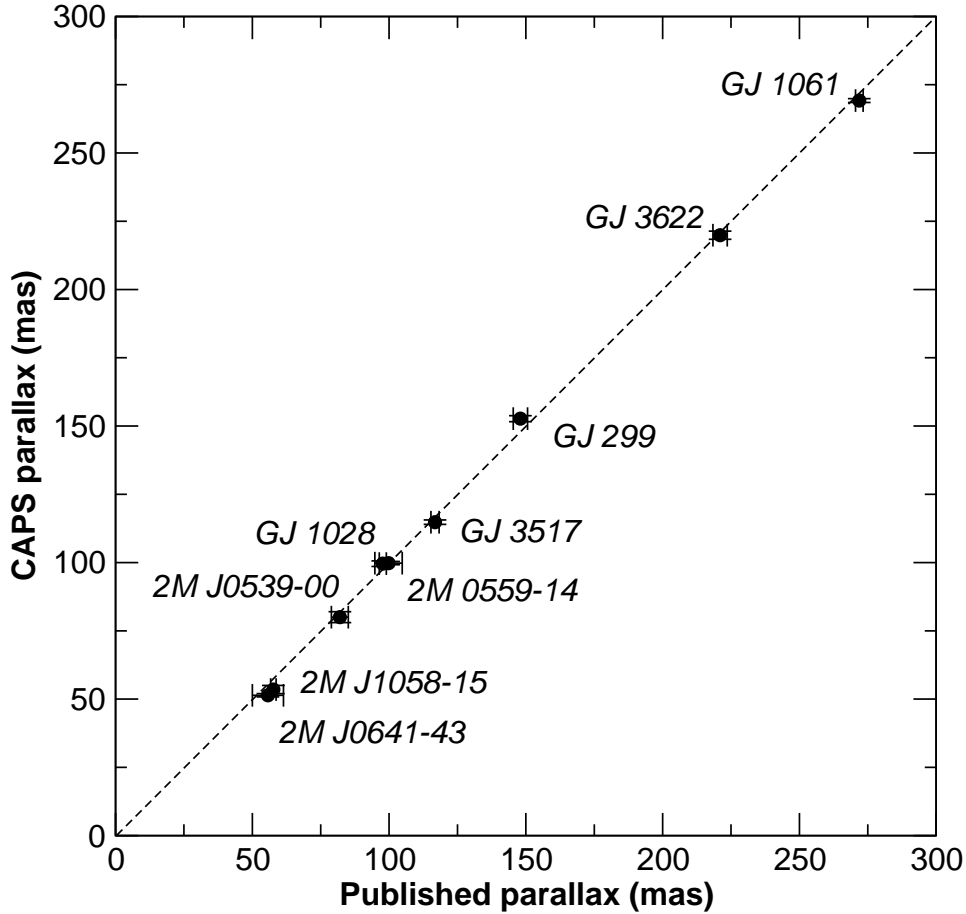}
\caption{Comparison of the previously published parallaxes to a
the CAPS program targets with more than 10 epochs. The published
parallaxes have been obtained from
SIMBAD (\texttt{http://simbad.u-strasbg.fr/simbad/}) and
references therein.
}
\label{fig:comparison}

\end{figure}

The measured motion of GJ 317 is relative to the background
stars. As discussed in Appendix \ref{sec:processing}, the
reference stars are matched to their predicted positions that, in
turn, are also refined on each iteration of the astrometric
iterative solution. As all the stars have the same parallactic
motion (except for the amplitude), the average parallactic motion
of the reference frame cannot be derived from relative
astrometric observations.  This translates into a zero--point
ambiguity in the measured distance to the target that needs to be
corrected. After the final solution is obtained for all the stars
in the field, the measured differential parallax of some of the
references is used to estimate the zero--point correction as
follows. Using the B, J, H and K magnitudes from the NOMAD
catalog \citep{nomad}, photometric distances are obtained to 7
stars with all these colors available. The information for all
the reference stars and is given in Appendix
\ref{sec:processing}. Photometric distances to main sequence
stars estimated this way can have uncertainties of up to 20\% of
their actual values. Note, however, that a star at 330 pc will
have a parallax of 3 mas and a corresponding uncertainty in the
photometric parallax of only 0.6 mas. In comparison, a star at 50
pc would have a parallax of 20 mas and a corresponding
uncertainty of 4 mas. Therefore, we only use stars with nominal
photometric distances beyond 300 pc to obtain a more secure
determination of the zero--point. The average difference between
the photometric parallaxes and the measured ones is the desired
offset and amounts to 0.23 $\pm$ 0.1 mas for this field in
particular. This offset is added to the measured parallax of GJ
317 to provide the final distance determination in Table
\ref{tab:star}. The fact that the dispersion around the average
zero-point is small, indicates that this strategy is robust
against uncertainties in the photometry and the models. To prove
that our procedure is essentially correct, Figure
\ref{fig:comparison} shows the parallaxes of several CAPS program
stars compared to the ones published in the literature. We
actually found several outliers (not shown here) that typically
correspond to previous parallactic measurements done with a small
number of epochs and a formal uncertainties larger than 5 mas.
These cases and an overview of the CAPS sample will be discussed
in a future publication. For the stars shown here, the
measurements coincide within the published uncertainties and we
find no systematic zero-point offset.

Our measurement of GJ 317's proper motion is similar to some
previous estimates \citep[e.g.][]{salim:2003}, which is another
consistency check of the astrometric calibration procedure (e.g.,
plate scale and absolute field rotation).  However, such proper
motion is also relative to the references and requires more
assumptions than the parallax zero-point to be properly
corrected. For galactic kinematic studies, the catalog values of
the proper motion given in \citet[][NLTT]{salim:2003} or
\citet[][UCAC3]{ucac3} should be used instead of our relative
measurement (see Table \ref{tab:star}). Our relative proper
motions in R.A. and Dec. agree within 15 and 2 mas yr$^{-1}$ to
the values given in both these catalogs. Even though they differ
by more than 1-$\sigma$ (statistical uncertainty), a discrepancy
at the level of 15 mas yr$^{-1}$ is still acceptable due to the
unknown zero-point in our relative measurements and the
uncertainty in the catalogs themselves (random errors but also
zonal systematic effects). GJ 317's proper motion is also given
in other catalogs; e.g. \citet[PPMXL][]{ppmxl} and
\citet[USNO-B1.0][]{usnob1}. We find that those measurements are
unreliable. For example, the values given in PPMXL differ by 560
and 1350 mas yr$^{-1}$ from our measurement. A similar
discrepancy is observed in USNO-B1.0, which is expected given
that PPMXL is derived from USNO-B1.0. Given that the field is
moderately crowded, we suspect that such mismatch is due to a
crossmatching error (two stars with large proper motions are
reported within 10$^{\arcsec}$ of the nominal position of GJ 317
in both catalogs). Even though GJ 317 is bright in the nIR, it
can be easily confused with background objects at visible
wavelenghts.

In summary, the measured trigonometric parallax of GJ 317 is
65.3$\pm$ 0.5 mas. This puts the star at a distance of 15.3 $\pm$ 0.2
pc compared to the value used by JB07 ($110\pm 20$ mas or 9.1$\pm$
1.6 pc) using \citet{jenkins:1963} and \citet{gliese:1991}). The
other most recent measurement is 94.2 $\pm$ 12 mas \citep{yale:1995},
still off by more than 2-$\sigma$ illustrating that previous
parallactic measurements with uncertainties larger than 10 mas have
to be taken with caution. Unfortunately, the astrometric amplitude
due to an unseen companion is inversely proportional to the distance
to the Sun. This effectively suppresses the expected astrometric
signal by a factor of 1.7, providing a minimum (edge-on)
semiamplitude of only 0.30 mas (compared to the 0.58 mas initially
expected). The analysis and the actual limits imposed by the
astrometry are discussed in Section  \ref{sec:analysis}.

\begin{deluxetable}{lrrr}
\tablecaption{Properties of GJ 317.
\label{tab:star}}
\tablewidth{0pt}
\tablehead{
\colhead{Parameter} & \colhead{Value} 
}
\startdata
\hline
$\mu_{R.A.}$ (mas yr$^{-1}$) & -438 $\pm 5$ \tablenotemark{a}\\
$\mu_{Dec.}$ (mas yr$^{-1}$) &  794 $\pm 5$ \tablenotemark{a} \\
Absolute parallax (mas)      & 65.3 $\pm$ 0.4 \tablenotemark{(*)}\\
$d$ (pc) & $15.10 \pm 0.22$  \tablenotemark{(*)}    \\
\\
Calan-ESO spectral type & M2.5V \tablenotemark{b}\\
$\left[ \rm{Fe/H} \right]$ & 0.36 $\pm$0.2 \tablenotemark{(*)}\\
$T_{\rm eff}$ (K) & $3510 \pm 50$ \tablenotemark{(*)}\\
                  & (fit to Baraffe+ 1998)\\
Adopted $M_*$ (M$_\sun$) & $0.42 \pm 0.05$\tablenotemark{c} \\
Heliocentric RV (km s$^{-1}$) & 87.8 $\pm$ 1.5 \tablenotemark{d}\\
Heliocentric UVW (km/s) & (-92.7, -52.5, 26.5)\tablenotemark{(*)} \\
Galactic UVW (km/s) & (-82.3, 178.3, 33.8)\tablenotemark{(*)} \\
\enddata
\tablenotetext{*}{This work.}
\tablenotetext{a}{Absolute proper motions from the NLTT catalog \citet{salim:2003}}
\tablenotetext{b}{\citet{rojo:2003}}
\tablenotetext{c}{Mass obtained using absolute magnitudes in J, H,
and K; and the \citet{delfosse:2000} calibration}
\tablenotetext{d}{\citet{gizis:2002}}
\end{deluxetable}

\subsection{Observations : Radial velocities}\label{sec:rv}

The 37 RV measurements presented herein were obtained with the HIRES
spectrometer \citep{vogt:1994} of the Keck I telescope from January
2000 to March 2010. This adds 18 measurements to those presented in
the discovery paper of GJ 317b/c \citep{johnson:2007}. Typical
exposure times on this star are 500 sec. The Doppler shifts are
measured by placing an iodine absorption cell ahead of the
spectrometer slit in the converging f/15 beam of the telescope. This
gaseous absorption cell superimposes a rich forest of iodine lines on
the stellar spectrum, providing a wavelength calibration and proxy
for the point-spread function (PSF) of the spectrometer. For the Keck
planet search program, we operate the HIRES spectrometer at a
spectral resolving power R = 70,000 and wavelength range of 3700-8000
A, though only the region 5000-6200 A (with iodine lines) is used in
the present Doppler analysis. Doppler shifts from the spectra are
determined with the spectral synthesis technique described by
\citet{butler:1996}. Due to several improvements in our RV extraction
pipeline for M-dwarfs \citep[see][for further details]{vogt:2010},
the new RV measurements update those given in \citet{johnson:2007}.
The HIRES/Keck has a demonstrated long term stability better than 3 m
s$^{-1}$ for other stars with similar spectral types
\citep[e.g.,][]{vogt:2010}. The analysis of the RV data is given in
Section \ref{sec:rvanalysis}. The differential radial velocity
measurements of GJ 317 are given in Table \ref{tab:rvdata} and the
average heliocentric radial velocity is given in Table
\ref{tab:star}. While the astrometry does not show any obvious
signal, GJ 317b can be clearly seen in the RV data. The signal of GJ
317c is seen as a long term quadratic drift in the Doppler
measurements (see Section \ref{sec:rvanalysis}).

\begin{deluxetable}{lcc}
\tablecaption{Differential radial velocity measurements of GJ 317
corrected by the barycentric motion of the observer\label{tab:rvdata}
}
\tablehead{
\colhead{Julian date} & 
\colhead{RV} & 
\colhead{Uncertainty}\\
\colhead{(days)} & 
\colhead{(m s$^{-1}$)} & 
\colhead{(m s$^{-1}$)}
}
\tablewidth{0pt}
\startdata
2451550.99259  &    7.34 & 3.73  \\
2451552.98958  &   21.52 & 5.07  \\
2451582.89062  &   49.03 & 4.05  \\
2451883.10117  &  -24.37 & 3.70  \\
2451973.79513  & -68.35  & 6.28  \\
2452243.07296  &   0.00  & 6.64  \\
2452362.94880  &  84.76  & 6.20  \\
2452601.04502  & -39.89  & 5.19  \\
2452989.12482  & 102.30  & 5.21  \\
2453369.01627  & -37.70  & 3.55  \\
2453753.98281  & 128.09  & 3.45  \\
2454084.00126  & -19.25  & 4.65  \\
2454086.14134  & -24.09  & 4.75  \\
2454130.08245  & -16.49  & 5.35  \\
2454131.01418  & -12.56  & 4.71  \\
2454138.93219  & -15.19  & 3.01  \\
2454216.73300  &   1.08  & 4.20  \\
2454255.74588  &  18.98  & 2.41  \\
2454400.10995  & 117.38  & 3.93  \\
2454428.06242  & 128.31  & 4.59  \\
2454464.99786  & 116.58  & 4.27  \\
2454490.95016  & 119.13  & 5.09  \\
2454492.90113  & 115.61  & 4.11  \\
2454543.94829  &  90.23  & 4.87  \\
2454544.90468  &  95.00  & 3.52  \\
2454545.89435  &  95.14  & 3.79  \\
2454600.82150  &  54.64  & 4.75  \\
2454806.02892  & -13.08  & 3.99  \\
2454807.06907  & -16.41  & 6.62  \\ 
2454808.13819  & -17.69  & 3.53  \\
2454809.05881  & -25.66  & 3.48  \\
2454810.16094  & -28.96  & 3.75  \\
2454811.12802  & -23.48  & 3.61  \\
2454820.95173  & -12.90  & 2.78  \\
2454822.98435  &  -8.78  & 2.77  \\
2454839.10383  & -29.39  & 4.46  \\
2455258.83718  &  92.12  & 4.74  \\
\enddata
\end{deluxetable}

\section{GJ 317 is a metal rich M dwarf}

GJ 317 (LHS 2037) was classified as an M3.5V star based on similarities
to the high resolution spectrum of the M3.5V M dwarf GJ 849 (JB07). The
assumed distance was 9.1 pc. In JB07, it was noted that the photometry
looked anomalous in the sense that the magnitudes did not match those
expected for a main sequence star. The comparison star(GJ  849) has a
HIPPARCOS parallax \citep{vanleeuwen:2007} also giving a distance of 9.1
$\pm$ 0.2 pc. Therefore, if both stars were of the same spectral type,
they should exhibit similar magnitudes. This is clearly not the case (on
average, GJ 317 is 1.5 magnitudes fainter in all the bands as given by
the
\textit{Simbad}\footnote{\texttt{http://simbad.u-strasbg.fr/simbad/}})
compilation. \citet{rojo:2003} used photometry and low resolution
spectroscopy to obtain a spectral type of M2.5V, but  reported a
photometric distance estimate of 7 pc, which puts the star even closer.
As we show in this section, the updated distance (15.3 pc) and, high
metallicity seem to solve most of these apparent contradictions.

\begin{deluxetable}{lrrr}
\tablecaption{Photometry of GJ 317. Absolute magnitudes assume
a distance of 15.3 $\pm$ 0.2 pc.
\label{tab:photometry}}
\tablewidth{0pt}
\tablehead{
\colhead{Band} & \colhead{Measured} & Absolute 
}
\startdata
B & 13.48 $\pm$ 0.03 \tablenotemark{a}  &  12.56 $\pm$ 0.04 \\
V & 11.98 $\pm$ 0.03 \tablenotemark{a}  &  11.06 $\pm$ 0.04 \\
R & 10.82  $\pm$ 0.03 \tablenotemark{a}&   9.89 $\pm$ 0.04 \\
I &  9.32  $\pm$ 0.03 \tablenotemark{a}&   8.39 $\pm$ 0.04 \\
J &  7.934 $\pm$ 0.03 \tablenotemark{b}&   7.01 $\pm$ 0.03 \\
H &  7.312 $\pm$ 0.07 \tablenotemark{b}&   6.39 $\pm$ 0.08 \\
K$_{s}$ &  7.028 $\pm$ 0.02 \tablenotemark{b}&   6.10 $\pm$ 0.03 \\
\enddata
\tablenotetext{a}{\citet{rojo:2003}}
\tablenotetext{b}{\,2MASS catalog, \citet{2MASS}}
\end{deluxetable}

\begin{figure}[tb]
\center
\includegraphics[angle=0, width=3in, clip]{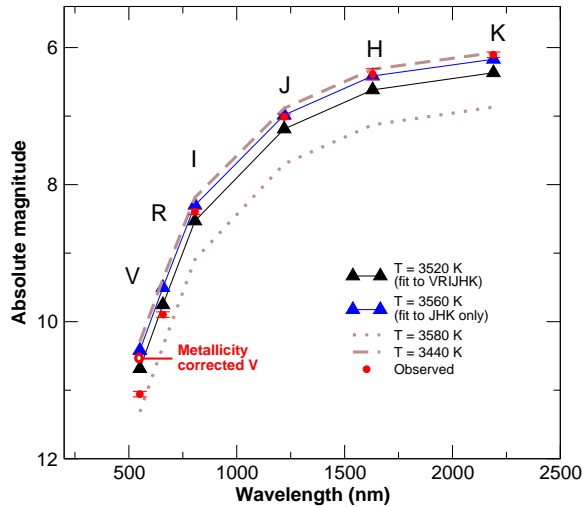}

\caption{Fit of the VRIJHK magnitudes to the \citet{baraffe:1998}
theoretical models (black triangles). Compared to the models, the
Visual V mangitude is fainter and the J,H,K colors appear brighter
than expected by many standard deviations. A fit only to the JHK
magnitudes is shown in blue triangles. A much better agreement is
recovered when the V magnitude is 'corrected' by its metallicity
(see text).
}\label{fig:baraffe}

\end{figure}

Given a trigonometric distance, the mass and temperature of GJ 317 can
be obtained by fitting the absolute magnitudes to the theoretical models
given by \citet{baraffe:1998}. Table \ref{tab:photometry} shows the
available photometry of GJ 317 (B, V, R, I, J, H and K$_s$) and the
updated absolute magnitudes. The adopted photometric measurements in the
visible are those given by \citet{rojo:2003}. The infrared magnitudes
have been taken from the 2MASS catalog \citep{2MASS}. Even though there
are other photometric measurements in the literature, most of them agree
within the reported uncertainties. The fit is obtained by adjusting the
absolute magnitudes to the values tabulated in \citet{baraffe:1998},
using the temperature as the only free parameter. Because the star has
low activity levels and low $v \sin i$ (JB07), we assume an age of 5 Gyr
(M dwarfs change their colors very little after 1 Gyr). This fit leads
to a preliminary mass of $M=0.38$ M$_\sun$ and an effective temperature
of T$_{\rm eff}$ = 3510 K. If only infrared colors are used (J, H and
K), a slightly hotter and more massive star is recovered; T$_{\rm eff}$
= 3550 K and 0.43 M$_{\sun}$. As can be seen in Fig. \ref{fig:baraffe},
the predicted flux in the optical wavelength range is still
overestimated by the models; that is, the star seems to be fainter than
expected in the bluer colors. While moderate interstellar absorption by
dust could be the cause, GJ 317 is a relatively nearby star and there is
no reported evidence of strong extinction in its direction. A$_V$ is 0.3
mag for the whole integrated galaxy
\footnote{\texttt{http://ned.ipac.caltech.edu/}} in the direction of GJ
317, and still would be insufficient to explain the observed excess. The
absolute magnitudes can also be used to infer the mass of GJ 317 using
the empirical relations given by \citet{delfosse:2000}. From the
absolute J, H, and K$_s$ magnitudes we obtain : $M_{*J} = 0.41 M_\sun$,
$M_{*H} = 0.42 M_\sun$, $M_{*K} = 0.43 M_\sun$, while using V and V-K,
we derive $M_{*V} = 0.37 M_\sun$ and $M_{*V-K} = 0.25 M_\sun$. These
mass estimates show a similar trend as the fit to the theoretical
models: longer wavelength photometry favor larger masses. As advised by
\citet{delfosse:2000}, we  adopt the average of the masses obtained from
the JHK magnitudes ($M_{*}=0.42 M_{\sun}$) as the most reliable
estimate. This value is 1.75 times larger than the one adopted by JB07
in the discovery paper of GJ 317b and c. Therefore, the derived minimum
masses for both planets will be increased accordingly (see Section
\ref{sec:rvanalysis}).

Metallicity has an effect on the colors of M dwarfs: a star with
super-solar metallicity will have the V-band flux suppressed compared to
a star of the same spectral type with solar metallicity. This effect has
been used to estimate metallicities of field M dwarfs, so it would not
be surprising that the visual mangitudes appear fainter than expected if
GJ 317 were, in fact, metal rich. The best calibrations use the V-K
color against the absolute K magnitude to estimate [Fe/H] :
\citet{bonfils:2005} or B05, \citet{johnson:2009} or JA09, and
\citet{schlaufman:2010} or SL10. All three calibrations are empirical
and are based on measured metallicities of nearby stars with good
parallax measurements. Based on the previous distance, JB07 obtained a
[Fe/H] = -0.23. This value made GJ 317 the only 'metal poor' M dwarf
with detected giant planet candidates. Given the updated M$_K$, all
three methods now indicate that the star is indeed metal rich ([Fe/H]=
+0.08 for B05, [Fe/H] = +0.43 using JA09, and [Fe/H] = +0.29 using
SL10). Since B05 tends to underestimate metallicities
\citep{johnson:2009,rojas:2010}, we adopt the average of the JA09 and
SL10 calibration, i.e. 0.36, as the updated value for [Fe/H].

We now investigate whether or not the color/metallicity relation can
explain the apparent extinction in the visible magnitudes when compared
to the models or to the mass/luminosity relations. Since the highest
metallicity comes from the JA09 calibration, and assuming that $M_{\rm
K}$ is correct, we ask what V magnitude would be required to conclude
that the star has [Fe/H] = 0.0 (this is the metallicity assumed by the
\citet{baraffe:1998} models). By iterating on the JA09 relation we find
that V=11.45 (compared to an observed V=11.98) gives zero metallicity. We
can now test whether this value can do a better job fitting the
theoretical models. We find that the correction is in the right direction
(see Fig. \ref{fig:baraffe}) and it significantly improves the fit to the
V, J, H and K photometry. Also, the revised mass derived from the models
is now 0.42 $M_\sun$, much closer to the one fitting the J, H and K bands
only ($0.43 M_\sun$). Moreover, if we apply the mass/luminosity relations
in \citet{delfosse:2000} to the 'corrected' V magnitude, we obtain
$M_{*V} = 0.43 M_\sun$ and $M_{*V-K} = 0.41 M_\sun$, now in perfect
agreement with the value obtained from J, H and K. A similar behaviour
would be expected for the R and I bands, but no empirical color
metallicity relations have been published on these bands.
\citet{delfosse:2000} already mentioned that the mass/luminosity relation
using the V color has an excess of dispersion due to the unknown
metallicities of M dwarfs. The remarkable agreement in all the quantities
recovered after 'correcting' V for the effect of metallicity seems to
confirm this. As a more definitive proof, it would be desirable to obtain
a direct metallicity measurement using the newly developed spectroscopic
methods in the near infrared \citep[e.g.,][]{rojas:2010}.

Given the new distance measurement, we can re-compute its tangential and
galactic 3D velocity to check whether GJ 317 could be a member of a
known kinematic group. The UVW velocities in the Heliocentric and the
Galactic reference frames are given in Table \ref{tab:star}. The
resulting U and V components are large but not unusual for disk stars.
We have integrated the Galactic orbit using a basic potential for the
Galaxy (thin disk, thick disk, bulge and halo, see \citet{pac:1991}) and
a 4th order Runge-Kutta integrator (time steps of 100 years). We find
that past and future trajectories lie well within the galactic disk
region. Actually, a large U component with a V component of the order of
$\sim -50$ km/s and a moderate W are typical of a thick disk population
member \citep[see Fig. 1 in][]{bensby:2007}. This would be consistent
with the star being old ($>3$ Gyr) and inactive. Even though the
boundaries between thick and thin disks are diffuse \citep{bensby:2007},
everything indicates that it is a likely a member of the Galactic thick
disk population.

In summary, the previously reported anomalies in the photometry of GJ 317
seem to be related to the previous poor estimate of its distance and its
high metal content. The new metallicity determination strengthens the
observed correlation of super solar metallicity and the presence of giant
planets around M dwarfs \citep{rojas:2010, johnson:2009}.

\section{Orbital analysis}\label{sec:analysis}

At the updated distance, our astrometric measurements cannot resolve the
wobble of planet b (minimum amplitude $\sim 0.3$ mas). Assuming an
epoch-to-epoch precision of 0.9 mas (average between R.A. and Dec. 
precisions) on 18 epochs and Gaussian statistics, an amplitude of 0.30
mas should be detected only with a signal-to-noise ratio of 1.4. An
amplitude of 0.65 mas should be detected with a SNR of 3 if present.
Therefore, at our present precision the amplitude of the signal can only
be marginally measured with a signal-to-noise ratio between 1 and 4
(depending on the actual orbital inclination). While this is
insufficient to obtain an unambiguous detection, our analysis shows that
we can actually put tight constraints on the mass of GJ 317b, thus
confirming its planetary nature. Because of the reduced sensitivity of
the astrometry, the orbital parameters constrained by the RV are not
sensitive to the astrometric measurements. Therefore, we first provide a
detailed analysis of the RV data and then we discuss the constraints
imposed by the astrometric measurements on GJ 317b and GJ 317c.

\subsection{RV analysis}\label{sec:rvanalysis}

We use the latest stable version of the \textit{systemic interface}
\citep[][v1.5.12]{systemic:2009} to obtain the best fit to the RV data.
GJ 317b is clearly seen in the periodogram as a strong peak around 700
days. After subtracting the best orbital fit for GJ 317b (see Table
\ref{tab:rvsolution}), a long term trend is still apparent in the
residuals. Assuming a circular orbit we obtain a period of 7100 days for
GJ 317c. Even in the circular orbit case, a broad range of periods and
masses are still allowed by the data. For an eccentric solution
($e_c\sim 0.81$), the most likely period falls between 50 000 and 90 000
days ($\sim$ 130--250 years). Compared to the orbital solution of JB07
based on 18 measurements, it is now more clear that a constant slope is
insufficient to reproduce the observed RVs and that the curvature of the
orbit is clearly detected in the data. In the case of eccentric orbits,
such long period signals can only be detected if the planet is close to
the periastron of its orbit, which might be the case here. In any case,
several more years of RV measurements will be required to put a more
significant constraint on this orbit. 

\begin{deluxetable}{lclc}
\tablecaption{Orbital solution for GJ 317b/c from the RV data
only. All parameters are referred to the initial epoch
$T_0$=2451550.9925. The mass of GJ 317 is assumed to be 0.42
M$_{\sun}$. 
\label{tab:rvsolution}}
\tablehead{
\colhead{} & 
\colhead{GJ 317b\tablenotemark{(+)}} & 
\colhead{GJ 317c} &
\colhead{GJ 317c} \\
\colhead{} & 
\colhead{ } & 
\colhead{(circular)} &
\colhead{(eccentric)} \\
}
\tablewidth{0pt}
\startdata
\\
$P$ (days)             &  692 $\pm$ 2    & 7100$^{+8000}_{-1500}$ & $>10^4$ \\
$K$ (m\,s$^{-1}$)      &   73.5$\pm$ 2   & 30.5	$\pm$ $5$ & $\sim$ 30    \\
$e$                    & 0.11 $\pm 0.05$ & 0 (fixed)  & 0.81 $\pm$ 0.2 \\
$M_0$                  & 309 $\pm$ 15    & 199$_{-25}^{+75}$& 350 \tablenotemark{(u)}    \\
$\omega$ (deg)         & 340 $\pm$ 10    & --         & 210 \tablenotemark{(u)} \\
$\gamma$ (m\,s$^{-1}$) &                 & 18         &  40 \tablenotemark{(u)} \\
\\
Derived quantities     &                 & 	      &	             \\
\hline\\
$M \sin$ i ($M_{jup}$) & 1.81 $\pm$ 0.05 & 1.6        & $\sim$2.0      \\
a (AU)                 & 1.148           & 5.5 AU     & 10-40 AU     \\
\\
Statistics\\
\hline\\
N$_{\rm RV}$          &                  & 37         & 37           \\
RMS (m\,s$^{-1}$)     &                  & 8.5        & 7.3          \\
$\chi^2_{RV}$         &                  & 4.97       & 4.36         \\
\enddata

\tablenotetext{(+)}{The solution for GJ 317b is quite insensitive to
the eccentricity of GJ 317c. This solution corresponds to the
eccentric fit for GJ 317c given on the rightmost column of this table.}
\tablenotetext{(u)}{ unconstrained}
\end{deluxetable}

We find that the orbital solution for GJ 317b remains well constrained
irrespective of the orbital fit to GJ 317c. We use a Bayesian Monte Carlo
Markov Chain (MCMC) approach to characterize the probability
distributions of the free parameters and get a realistic estimate of
their uncertainties. The methodology applied to obtain such distributions
is described in Appendix \ref{sec:bayes}. Using only RV observations, we
find that the orbit of planet b is well constrained in a narrow region of
the parameter space, while the period and eccentricity of planet c still
have broad probability distributions (the 99\% confidence level interval
allows 20 $<$ P $<$ 240 years if the orbit is allowed to be eccentric).
These broad distributions are caused because planet c is only detected as
a secular quadratic term that can be reproduced by many combinations of
parameters (see \citet{gould:2008} for a detailed discussion of the
problem). The best fit orbital solutions for GJ 317b and GJ 317c are
shown in Table \ref{tab:rvsolution}. The uncertainties were obtained
using the aforementioned MCMC technique and represent the 68\% confidence
level intervals around the least-squares solution. Note that, because the
mass of the star has been updated, the minimum mass of planet b (1.8
$M_{Jup}$) is larger than the value given by JB07 (1.2 $M_{Jup}$).

\begin{figure}[tb] 
\center 
\includegraphics[angle=0, width=5in,clip]{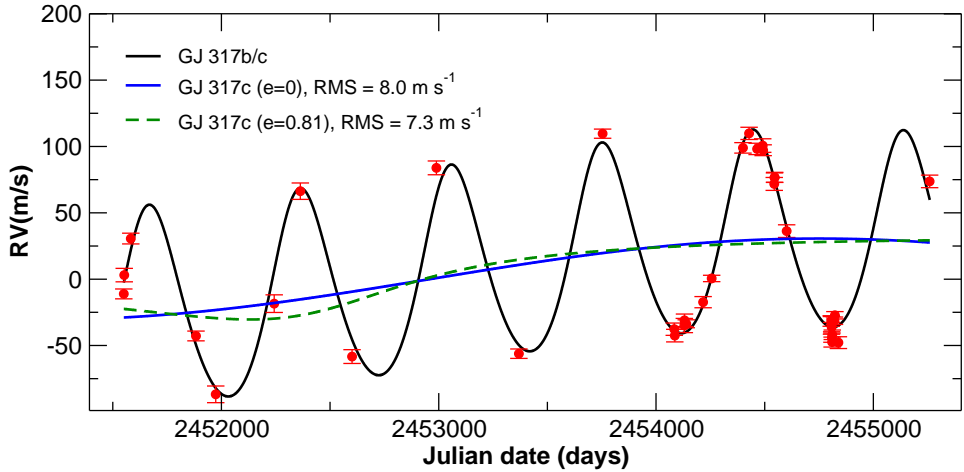} 
\caption{Best orbital fit to the RV data, assuming an eccentric orbit for
planet GJ 317c (solid black line). The signal of GJ 317c can be seen in
the secular trend of the RV signal of GJ 317b. We overplot the signal of
GJ 317c assuming  a circular orbit (solid blue) and the best eccentric
orbital fit (dashed green). \label{fig:rvfit}
}
\end{figure}

After removing both signals, there is no conclusive indication of
additional planets in the RV data. However, the root mean squared (RMS)
of the residuals ($\sim$8 m s$^{-1}$) is significantly larger than that
expected from observations of similar stars without planets (RMS $\sim$
3 m s$^{-1}$, see Section~\ref{sec:rv}). Also, the short period domain
(a few days to weeks time scales) is still poorly sampled. Given the
abundance of low mass objects in multi--planet systems
\citep{howard:2010}, additional low mass companions might be expected
to emerge as more RV measurements are obtained.

\subsection{Astrometry and radial velocities}
\label{sec:astroanalysis}

\begin{figure}[tb]
\center
\includegraphics[angle=0, width=5in, clip]{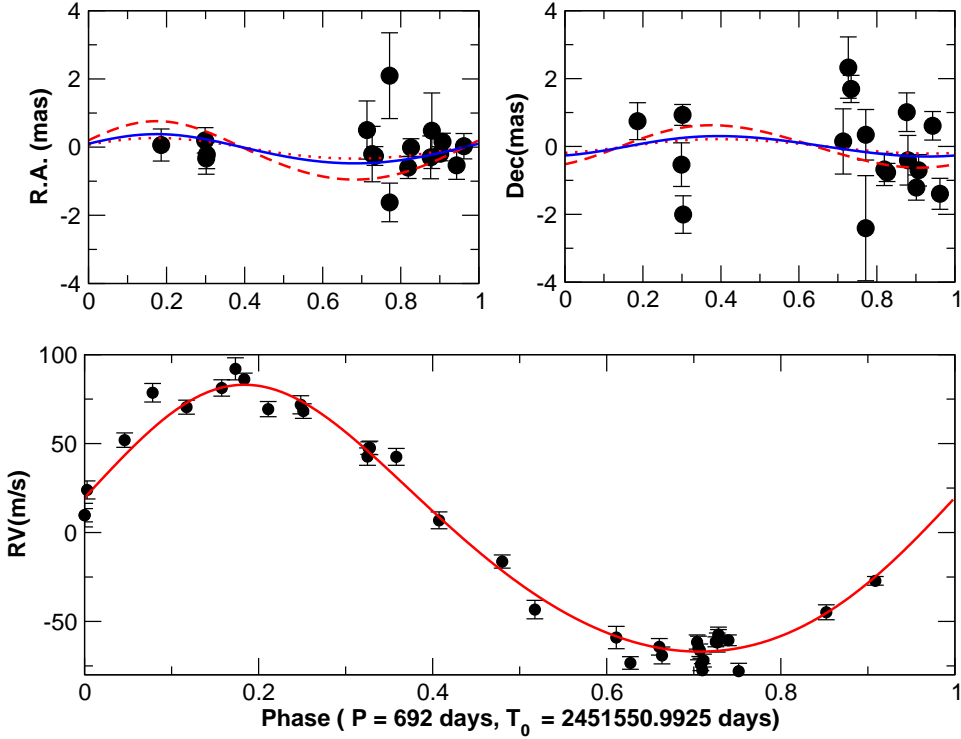}

\caption{Best fit astrometry + RV joint solution for GJ 317b. Top
panels are R.A. and Dec. as a function of time.  The dashed--red
line corresponds to the signal of a 4.0 $M_{Jup}$ mass
planet. The dotted--red line corresponds to the signal of a planet with
the minimum mass allowed by the radial velocities (1.8 $M_{Jup}$).
The best fit solution is plotted as a blue line. Bottom panel: the
radial velocity data plotted with the best fit solution after
removing the signal of planet c.}

\label{fig:astrorvsol}
\end{figure}

To perform a joint fit of the astrometry and the RVs, we subtract the
best orbital solution for planet c (e = 0.81), leaving only the RV signal
of GJ 317b in the RV data.  The least-squares solution for the combined
astrometry and radial velocities is obtained on a grid of fixed
period/eccentricities around 690 days (50 test periods between 680 days
and 700 days and 20 test eccentricities from 0 to 0.95). All the other
parameters are left free. In a last step, the period and the eccentricity
are refined starting at the best solution on the grid. The best fit
values (see Fig. \ref{fig:astrorvsol} and Table \ref{tab:astrorvsol}) are
then used to initialize a Bayesian Monte Carlo Markov Chain sampler to
generate again the \textit{a posteriori} probability distributions,
including now both the RVs and the astrometry in the definition of the
likelihood function. The condition equations that the astrometry and the
radial velocities have to satisfy together with a brief explanation of
the Bayesian MCMC method are given in Appendix \ref{sec:bayes}. The same
analysis procedure was also used in \citet{anglada:2010} to combine
astrometric and RV measurements and rule--out the existence of the
astrometric planet candidate VB10b, at least in moderately eccentric
orbits. As in the RV analysis, the steps in the Markov Chain sampler are
tuned to accept 15\% to 30\% of the proposed updates, and the first
10$^5$ steps are not used in the analysis to avoid oversampling the
favoured solution. Chains with 5$\times$10$^6$ steps are used to generate
the numeric realization of the \textit{a posteriori} distributions for
all the parameters. We repeat the process several times and compare the
resulting distributions to be sure that the chains are properly converged
obtaining good agreement in all the runs.

\begin{figure}[tb] 
\center 
\includegraphics[angle=0, width=3in,clip]{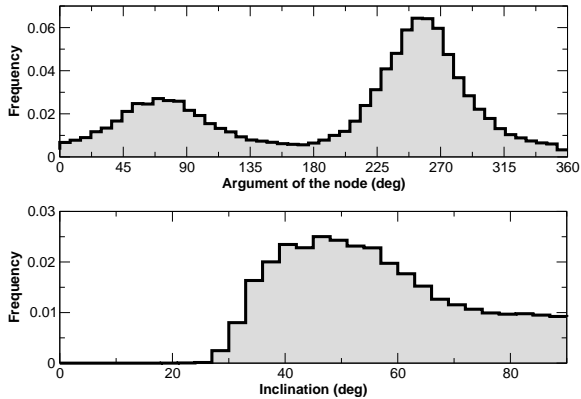}
\caption{Bayesian probability distributions for the orbital parameters
constrained by the astrometry observations. Because the signal is only
marginally detected, the argument of the node is poorly constraint. A
robust lower limit for the inclination is obtained.} 
\label{fig:OMEGAINC} 
\end{figure}

\begin{figure}[htb]
\center
\includegraphics[angle=0, width=3in, clip]{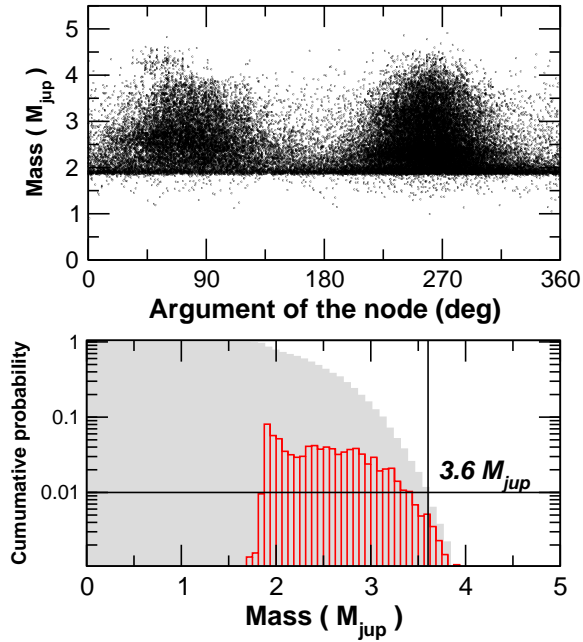}
\caption{Top. Markov Chain results illustrating the probability distribution of
the argument of the node of GJ 317b plotted against the mass (only one of
every 100 steps is shown to improve visualization). Bottom. In red we
show a histogram of the marginalized distribution for the mass of GJ
317b. The cumulative probability for the maximum mass of GJ 317b is
shown in grey. The RV data acts as a strong prior (abrupt cut below 1.8
$M_{Jup}$), effectively suppressing solutions below the minimum mass.
The upper limit on the mass is set by the astrometry at 3.6 $M_{jup}$.}
\label{fig:mass}
\end{figure}

The best fit solution (maximum likelihood values) with the corresponding
68\% confidence level intervals are given in Table \ref{tab:astrorvsol}.
Flat prior distributions have been used in all the cases. Figure
\ref{fig:OMEGAINC} shows the final probability distributions for the two
parameters that the astrometry can constrain. The fact that the signal is
barely detectable bodes ill for the determination of the argument of the
node (upper panel). However, with a 99\% confidence level, the
inclination is constrained to be greater than 25 deg (see bottom panel in
Figure \ref{fig:OMEGAINC}) because a value lower than that would result
in a wobble that is not seen. The top panel in Figure \ref{fig:mass}
illustrates the distributions for the mass and the argument of the node.
The distribution for the mass of GJ 317b is clearly non-Gaussian due to
the strong lower limit imposed by the RV data on the minimum mass and the
more loose constraints imposed by the astrometry on the maximum allowed
signal. The bottom panel of Figure \ref{fig:mass} shows the histogram of
the probability distribution for the mass. Note that an inclination
closer to 0 would have two effects : first, the amplitude grows as
$1/\sin i$, and second, as the orbit becomes face on, the amplitude
becomes more '2-dimensional'. This is, an edge-on orbit would be seen as
a line, while a face-on orbit (inclination close to 0) would show up as a
circle increasing its statistical significance. To obtain the range of
allowed masses, the obtained MCMC distribution can be numerically
integrated from $M$ to $\infty$ (shaded histogram in bottom panel of
Figure \ref{fig:mass}). We find that the probability of $M> 3.6 M_{jup}$
is only 1\%. This value is much lower than 13 $M_{\rm jup}$ (approximate
planet/brown dwarf boundary) and unambiguously confirms that GJ 317b is
an actual planet.

\begin{deluxetable}{lclc}
\tablecaption{Best orbital Solution for GJ 317\,b. All parameters are
referred to T$_0$=2451550.9925.
\label{tab:astrorvsol}}
\tablehead{
\colhead{RV observables} & &\colhead{Astrometric observables} &
}
\tablewidth{0pt}
\startdata
$P$ (days) &  692 $\pm$ 2             & Relative $\mu_{R.A.}$ (mas/yr$^{-1}$) & -457.8 $\pm$ 0.5 \\
$K$ (m\,s$^{-1}$) &  75.2 $\pm$ 3.0   & Relative $\mu_{Dec.}$ (mas/yr$^{-1}$) & 796.5 $\pm$ 0.5 \\
$e$ & 0.11 $\pm$ 0.05                 & Relative parallax (mas) & 65.0 $\pm$ 0.4 \\
$M_0$ (deg) & 305 $\pm$ 15            & $\Omega$ (deg) & 82 \tablenotemark{(u)} \\
$\omega$ (deg) & 342 $\pm$ 10         & $i$ (deg) & 45 $^{+30}_{-10}$  \\
\\
Statistics & & Derived quantities &\\
\hline\\
N$_{\rm RV}$ & 37                   & $P$ (years) &  1.894 $\pm$ 0.013 \\
N$_{\rm astro}$ & 17 $\times$ 2     & $M$ $\sin i$ (M$_{jup}$) & 1.8 $\pm$ 0.05 \\
RMS$_{\rm RV}$ (m\,s$^{-1}$) & 7.4  & $M$ (M$_{jup}$)	    & 2.5$^{+0.7}_{-0.4}$\\
RMS$_{\rm R.A.}$ (mas) & 0.70       & $a$ (AU) & 1.15 $\pm$ 0.05 \\
RMS$_{\rm Dec.}$ (mas) & 1.23       & Angular separation (mas)& 76\\
T$_P$ (Julian date) &  2451656.7\tablenotemark{a} $\pm$ 35 \\
$\chi^2_{RV}$/N$_{\rm RV}$  & 3.1                  \\
$\chi^2_{R.A.}$/N$_{\rm astro}$ & 1.1                \\
$\chi^2_{Dec.}$/N$_{\rm astro}$ & 4.8                 \\
\enddata
\tablenotetext{a}{Time of Passage through the periastron closer to
the initial epoch T$_0$.}
\tablenotetext{(u)}{Unconstrained}
\end{deluxetable}

Bayesian methods are very sensitive to the correct estimation of the
single measurement uncertainties, so we performed a few additional
tests to check the robustness of the upper limit found for the mass.
First, we repeat the whole process (least-squares grid + MCMC)
assigning a constant 1 mas uncertainty to weight each measurement. The
actual maximum mass at the 99\% confidence level (C.L.) is a bit larger
(3.8 $M_{jup}$). This comes from the fact that we are effectively
downgrading the confidence in the R.A. measurements, which are the more
constraining ones. We also tried shuffling the residuals in the
astrometry and then repeating the orbit fitting process on the shuffled
sets. We find a slightly larger upper limit on the mass of the shuffled
sets ($\sim 4 M_{Jup}$) than on the real data, confirming that the
signal is just barely resolved by the astrometry, probably only in R.A.
The bias towards low inclinations (large dynamical masses) in the
presence of noise is a known effect in the analysis of astrometric
observations \citep{pourbaix:2001}. Therefore, the upper limit we find
to the mass of GJ 317b is a very robust result.

\begin{figure}[htb]
\center
\includegraphics[angle=0, width=4in, clip]{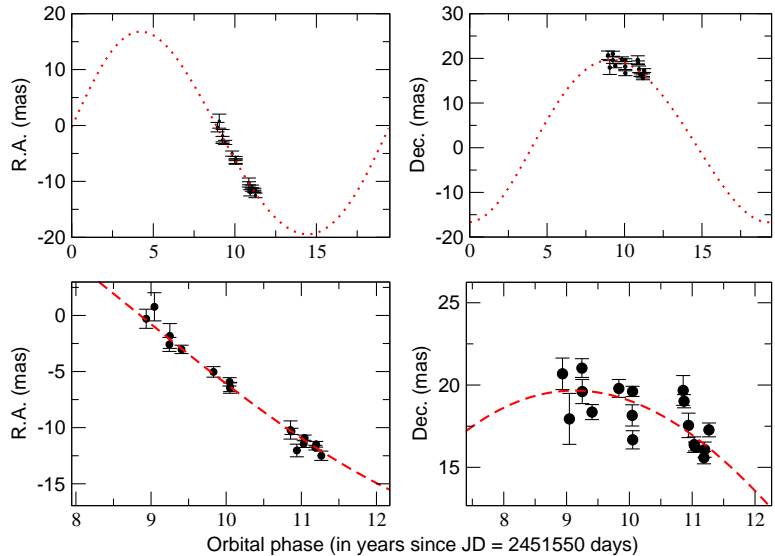}
\caption{Top panels : Best orbital solution using the astrometry on
planet GJ 317c. Bottom panels : Zoom-in on the orbit as sampled
by the CAPScam astrometry. Because it has a very long period,
the signal is mostly a linear trend that cannot be de-coupled from
the proper motion yet \citep{black:1982}. In 3 to 4 years, the
acceleration should be obvious in (at least) one of the directions.
}
\label{fig:longperiod}
\end{figure}

We also tested if the astrometry could put constraints on the
inclination and the mass of the outer planet c. To do this, we removed
GJ 317b from the radial velocities and combined the astrometry and the
radial velocities in a Bayesian MCMC initialized at the best circular
orbit from the RV. This time, because the period is so long, the
eccentricity is poorly constrained and because there is no appreciable
acceleration in the astrometry (see the residuals in Fig.
\ref{fig:rawastrometry}), the true mass of the companion is still very
uncertain. We note that if a linear trend in the astrometry were large
enough to be detected, then this trend would be absorbed by the fit to
the proper motion. This is a known issue for astrometric planet
detection and was first discussed by \citet{black:1982}. Assuming a
circular orbit, masses up to 22 $M_{Jup}$ are still compatible with the
astrometry (at a 99\% C.L.) If the orbit is allowed to be eccentric,
masses up to 200 $M_{Jup}$ are still allowed on orbits close to
face-on. We anticipate that, at least, 3-4 more years of follow up will
be necessary to obtain a first hint of the astrometric acceleration
(see Figure \ref{fig:longperiod}) and put a meaningful constraint on
the mass of GJ~317c. Depending on the actual age  of the system, direct
imaging of GJ 317c might be achievable with state-of-the-art high
contrast imaging systems\citep[e.g.,][]{lagrange:2010}. Imaging
combined with RV and astrometric measurements will provide a direct
measurement of the mass of the planet and the mass of the star.

\section{Discussion}

We have obtained precision astrometric measurements of the M dwarf GJ
317 and combined these with new radial velocities. Even though the
signal is not fully resolved by the astrometry, the new measurements
put meaningful constraints on the mass of GJ 317b, confirming its
planetary nature. Combining astrometric and radial velocity
measurements is a complex multi-parametric problem where the final
probability distributions for the involved parameters are not
necessarily Gaussian. This upper limit has been obtained using a
Bayesian MCMC approach which is much better suited than a classical
$\chi^2$ analysis to put contrains on parameters and obtain
confidence intervals. Given that the upper limit to the mass of GJ
371b ($\sim 3.6 M_{jup}$) is much lower than the planet/brown dwarf
boundary, it is the first time ground-based astrometric observations
have been able to confirm the planetary nature of a substellar
companion. The presented radial velocities confirm the presence of an
extremely long period planet (period 20 years or more) that is not
yet detected in the astrometry. 

Other RV surveys \citep[e.g.,][]{endl:2009,johnson:2007} have found a low
occurrence rate of moderate--to--short period gas giants around M dwarfs,
with the resonant pair GJ 876b/c the only remarkable exception
\citep{rivera:2010}. No hot Jupiters (P$<$30 days) have been  reported
around a low mass star. A handful of gas giants with periods longer than
30 days have been found around a few early type M dwarfs (e.g. GJ~179b, 
GJ~832b, GJ~849b, HIP~79431b, HIP~57050b, see the \textit{exoplanet
encyclopedia} for an up-to-date list\footnote{http://exoplanet.eu}). This
seems to indicate that M dwarfs have  trouble forming and/or keeping
giant planets in tight orbits.

According to recent studies, the frequency of M dwarfs hosting gas
giants seems rather low compared to more massive solar type stars
\citep[e.g.][]{johnson:2010}. This is an expected consequence of the
core accretion model for giant planet formation 
\citet{laughlin:2004,kennedy:2008,alibert:2011}. Also, the new distance
measurement indicates that GJ~317 is metal rich. With GJ~317 now in the
club, all the M dwarfs with reported giant planets are metal rich
\citep{rojas:2010,johnson:2009}. The competing mecanism model for planet
formation \citep[][disk instability model]{boss:1997} should not be very
sensitive to the metallicity of the host star, so this has also been
suggested as evidence in favor the core-accretion scenario
\citep[e.g.,see][]{ida:2004,mordasini:2009}. However, core accretion is
still too slow to form long period gas giant planets around low mass
stars, except for exceptionally long lived disks \citep{ida:2005}. Also,
core-accretion fails to reproduce the over-abundance of planets in
close-in orbits around all stellar types found by Kepler/NASA
\citep{borucki:2011}. On the other hand, disk instability is able to
form gas giants around M dwarfs with short lifetime disks
\citep{boss:2006,boss:2011}. Given these two issues (time-scale problem
and incorrect prediction of the observed planet distributions), we find
it more natural to invoke the disk instability mechanism to explain the
formation of GJ~317c, and possibly GJ~317b as well. Given that the
number of nearby M dwarfs surveyed for planets is still small, more
detections in a statistically larger sample are required to put real
constraints on the metallicty-gas giant connection. This is one of the
long term goals of the Carnegie Astrometric Planet Search project.

GJ~317 is one of the faintest targets in the Lick-Carnegie planet
search program, and precision radial velocities in the optical are
limited by photon noise and require intensive use of large aperture
telescopes (Keck/HIRES) to reach the few m/s precision level. Since it
is a cool star (T$<$4000 K), most of its flux is in the near infrared;
so it will be an excellent target for precision RV measurements when
the new generation of near infrared spectrographs comes on-line
\citep{bean:2010,figueira:2010,anglada:2011}. Also, the star lies in
the optimal magnitude range (12$<$V$<$15) for the Gaia/ESA space
astrometry mission \citep{gaia:2010} and the PRIMA/VLT interferometer
\citep{koehler:2010}. Both instruments will have a single measurement
precision better than 0.1 mas and the orbit of GJ 317b (minimal
semiamplitude of 0.3 mas) will be clearly resolved once an orbital
period ($\sim$ 700 days) is covered. Since the astrometric signal of
the long period planet GJ 317c should be evident in a few more years
of CAPScam observations as a quadratic term, we will continue
monitoring this star at lower cadence to measure its orbital
inclination and mass.

Finally, we highlight some unique features of GJ 317. It is a relatively
bright, low mass star with super solar metallicity, so it seems an ideal
target to look for low mass terrestrial planets in its habitable zone.
Recent studies \citep{mayor:2009,howard:2010} indicate that 30\% of the
dwarf stars host planets in the super-earth mass range. This fraction
seems to be even higher in multi-planetary systems. Such planets in the
habitable zone around GJ 317 should have orbital periods of a few weeks
and RV amplitudes of several m s$^{-1}$.  The fact that the system
contains a pair of long period giants might have helped to deliver
volatile compounds to the inner orbits as seemed to have happened in the
early Solar System \citep{crida:2009}. On the other hand, significantly
eccentric orbits are possible for both outer planets, which might be a
problem for the long term stability of the inner orbits
\citep{chambers:2002}. Finally, the outermost giant planet might be
detectable by direct imaging in the near future \citep[e.g.][found a sub
stellar companion to $\beta$ Pic at a similar angular
separation]{lagrange:2010}. An image of this planet combined with the
astrometry and the RV would provide a model independent measurement of
its mass. In summary, in a few years GJ 317 might become one of the
better characterized planetary systems beyond our own.

\textit{Acknowledgements.} G.A. gratefully acknowledges the support of
the Carnegie Postdoctoral Fellowship program. G.A. also thanks Cassy
Davison and Todd Henry, both at Georgia State Universtity, for valuable
comments and discussions. A.P.B., A.J.W. and G.A. gratefully acknowledge
the support provided by the NASA Astrobiology Institute grant NNA09DA81A,
and the continuous support of the Carnegie Observatories TAC to the CAPS
program. R.P.B. gratefully acknowledges support from NASA OSS Grant
NNX07AR40G, the NASA Keck PI program, and from the Carnegie Institution
of Washington. S.S.V. gratefully acknowledges support from NSF grant
AST-0307493. Some of the work herein is based on observations obtained at
the W. M. Keck Observatory, which is operated jointly by the University
of California and the California Institute of Technology, and we thank
the UC-Keck and NASA-Keck Time Assignment Committees for their support. 
The Lick--Carnegie group members acknowledge the contributions of fellow
members of our previous California Carnegie Exoplanet team in helping to
obtain some of the earlier RVs presented in this paper.

{\it Facilities:} \facility{Du Pont (CAPScam)}, \facility{Keck:I (HIRES)}


\appendix

\section{The ATPa pipeline : Astrometric extraction, 
calibration and solution}\label{sec:processing}

The reduction of the astrometric observations is done using the ATPa
software, specifically developed for the CAPS project \footnote{
Software available at :
\texttt{http://www.dtm.ciw.edu/users/anglada/Software.html}}. Further
details on the various steps of the astrometric data reduction are given
in \citet{boss:2009}. The astrometric processing done by ATPa can be
outlined as follows. First, the position of all the stars in each image
is extracted and mapped to a tanget plane to the sky centered at the
nominal coordinates of the target star at the first epoch of
observation. For each image, a subset of reference stars are matched to
their predicted position to adjust the telescope pointing, plate scale,
rotation and fit for a geometric field distortion. These predicted
positions are initialized using the stars extracted from a high quality
image. Using this image-by-image solution, the position of each star in
the field is mapped to the frame defined by the reference stars. Then,
all the measured positions in a night (or epoch) are averaged to obtain
the epoch position and its uncertainty of each star in the field.
Finally, all the epochs are fitted to a linear astrometric model
(position, proper motion and parallax) and the reference frame is
refined by selecting those stars showing the smallest residuals. This
process is iterated a few times untill convergence is reached, this is,
when the root-mean-squared (RMS) of the residuals of the reference stars
does not change significantly (e.g. less than 0.05 mas). We call this
process the \textit{astrometric iterative solution} (or AIT). The result
is a catalog of positions, parallaxes and proper motions for all the
stars in the field.

Now let's describe each step in more detail. The centroids of the
stars are measured by binning the stellar profile in the X and Y
directions on a box of 12 pixels ($\sim 2\arcsec$) around the pixel
with maximum flux. This one dimensional profile is then precisely
centroided using a function called \textit{Tukey's
biweight function}\citep[see p. 448 on][]{tukey}. Many simulations
and tests with different profiles showed that this approach provided
the best centroid accuracy and robustness \citep[see][for further
details]{boss:2009}. The flux is also measured on a circular
apperture of 10 pixels around the obtained centroid and the sky
obtained from a ring of 15 to 20 pixels is then subtracted. As a
result, each image produces a list of star positions and fluxes. A
preliminary centroid accuracy for a given night is empirically
obtained using a rough reference frame consisting on the the best 10
stars in the field in terms of centroid nominal precision. Finally,
the nominal optical distortion of CAPScam is used to map the pixel
positions to a local tangent planet to the sky. This nominal field
distortion was obtained in April 2011 by observing a moderately
crowded field on a rectangular grid of 16 positions spaced by
40$\arcsec$ (North--East tip of the rectangle centered at 16:00:11.41
-40:12:42.2). This method of measuring the nominal optical distortion
is described in \citet{anderson:2007} and references therein.

To initialize the astrometric iterative solution, the extracted
positions of one image are used as a template to generate an initial
catalog. In this first iteration, all the images are matched to this
catalog using a linear distortion model. This is, assuming that the
predicted local tangent plane position of the i-th reference star is
($u_i$, $v_i$), the geometric calibration consists in finding the
coefficients that satisfy

\begin{eqnarray}
u_{i} &=& a_0 + a_x x_{i} + a_y y_{i}\, ,\label{eq:distortion1} \\
v_{i} &=& b_0 + b_x x_{i} + b_y y_{i}\, ,\nonumber
\end{eqnarray}

\noindent where $x_{i}$ and $y_{i}$ are the extracted positions of
the references obtained in the ONP step after applying the nominal
CAPScam distortion. The same transformation is then applied to all
the stars in the field. The nightly averaged $u$ and $v$ of each star
is obtained (this is what we call an astrometric epoch measurement)
and the intranight standard deviation divided by the square root of
the number of observations is used as the associated epoch
uncertainty (i.e. uncertainties in Table \ref{tab:astrodata}).
Finally, all the epochs are used to update the astrometric solution
for each star by fitting their motion to

\begin{eqnarray}
u(t) &=& u_0 + \mu_{\alpha} \left(t-t_0\right) - \Pi p_{\alpha}(t) + \Delta_\alpha\, ,\label{eq:model}\\
v(t) &=& v_0 + \mu_{\delta} \left(t-t_0\right) - \Pi p_{\delta}(t)+ \Delta_\delta\, ,
\end{eqnarray}

\noindent where $\mu_{\alpha}$ and $\mu_{\delta}$ are the proper
motion in the direction of increasing R.A. and Declination
respectively (in mas yr$^{-1}$), $\Pi$ is the parallax in mas,
$u_0$ and $v_0$ are constant offsets (in mas) that provide the
star position on the local plane at the reference epoch $t_0$
assuming zero parallax. The numbers $p_{\alpha}$ and $p_\delta$
are the so-called parallax factors and correspond to the
parallactic apparent motion projected on the direction of
increasing R.A. and Dec. at the observing instant $t$. To ensure
maximal precision, the parallax factors are derived using the
position of the geocenter from the DE405 JPL Ephemeris of the
Solar System
\footnote{http://ssd.jpl.nasa.gov}. Note
that the 5 free parameters on this equation (u$_0$, v$_0$
,$\mu_\alpha$,$\mu_\delta$ and $\Pi$) are linear. Therefore, the
corresponding system of normal equations can be efficiently
solved in a single least-squares step. The perspective
acceleration for GJ 317 \citep{dravins:1999} is negligible given
the relatively short time baseline of this dataset ($\sim$ 0.042
mas over two years) and is not included in the processing. A
simple derivation of this astrometric model and additional second
order terms are outlined in \citet{anglada:2010b}.
$\Delta_\alpha$ and $\Delta_\delta$ contain possible
perturbations to the baseline astrometric model but they are not
adjusted during the AIT. For the target star, $\Delta_\alpha$ and
$\Delta_\delta$ will include the astrometric Keplerian motion
following the prescriptions outlined in Appendix
\ref{sec:bayes:astro}. Such functions implicitly depend on the
Keplerian parameters in a complicate non-linear fashion,
specially when combining astrometric and radial velocity
observations.

\begin{figure}[tb]
\center
\includegraphics[angle=0, width=5in, clip]{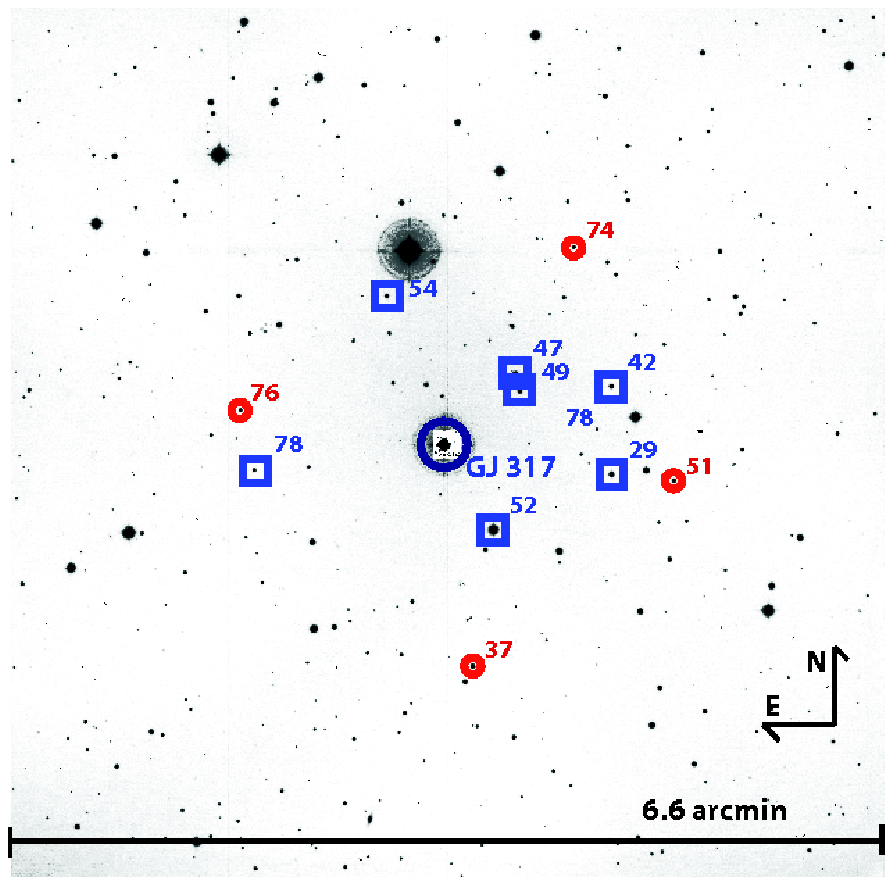}
\caption{GJ 317 field of view as seen by CAPScam on March 18th, 2011.
Reference  frame stars are marked with circles and squares. Stars
marked with blue squares are the ones used to derived the zero--point
parallax. Stars marked as red circles are still used as references
but could not be used to obtain the zero--point correction because
one or more photometric bands required (B, J, H or K) were missing in
NOMAD.} \label{fig:field}  
\end{figure}

Because this field is particularly rich in stars, our sotfware is
able to automatically select 11 very stable references within
120$^{''}$ of the target. By stable we mean that the RMS of the
epoch-to-epoch residuals (RMS) is smaller than 1 mas. In a second
iteration, these 11 references are used to recompute the field
distortion of each image with respect to the updated version of
the catalog. This time, a second order distortion correction (6
coefficients are adjusted on each axis) is fitted to each image.
This is,

\begin{eqnarray} 
u_{i} &=& a_0 + 
          a_x x_{i} + a_x y_{i} + a_{xx}x_i^2+
          a_{yy} y^2_{i} + a_{xy} xy\, ,\label{eq:distortion2} \\ 
v_{i} &=& b_0 +
          b_x x_{i} + 
	  b_x y_{i} + b_{xx}x_i^2+ b_{yy}y^2_{i}  + b_{xy} xy\, \nonumber\\ 
\end{eqnarray}

\noindent The RMS of each star and the epoch uncertainty derived
in the previous iteration are used to obtain a weighted solution
of Eq. \ref{eq:distortion2}. A 3-$\sigma$ clipping of the
outliers is also required to obtain a more robust determination
of the field distortion. Finally, an updated version of the
catalog is obtained by fitting Eq. \ref{eq:model} again to all
the stars. Because the precise positions and motion of the
references are not known a priori, the process has to be iterated
a few times. The convergence criteria is that the average RMS of
the references changes by less than 0.05 mas with respect to the
previous iteration. For this field in particular, only 5
iterations were required to reach convergence. The astrometric
solution for the reference stars and their catalog information
\citep{nomad} are given in Tables 7 and 8, and their
distribution on the field of view of CAPScam is illustrated in
Figure \ref{fig:field}.

\section{Bayesian Monte Carlo Markov Chains} \label{sec:bayes}

Bayesian statistics allow computation of the probability distributions
of the free parameters constrained by a set of observations. The general
strategy we use in this work is based on the methods described in full
detail in \citet{ford:2005}, which we strongly encorauge to read. To
obtain the parameter distributions compatible with the data we have to
obtain the likelihood function which, in this case, is just the
productory of the probability distributions of all the available
observations. This is, assuming that the uncertainties in in the
measurements follow a Gaussian distribution, the likelihood function $L$
reads

\begin{eqnarray} 
L &=& \kappa e^{-\frac{1}{2}\,\chi^2[\hat{\alpha}]}\,, 
\label{eq:likelihood} \\ 
\chi^2[\hat{\alpha}] &=& 
\sum_k^{N_{\rm obs}} 
\left(\frac{x_{\rm obs}-x_{\rm model}[\hat{\alpha}]}{\sigma_k}\right)^2 
\nonumber
\end{eqnarray}

\noindent where $\chi^2$ is the classic definition of the weighted
least-squares statistic, $x_{\rm obs}$ can be any kind of observations
(e.g. an RV measurement, an astrometric offset, an instant of transit),
$\sigma_k$ is the uncertainty on such measurement, $x_{\rm
model}[\hat{\alpha}]$ are the predictions of the model to be tested,
$\hat{\alpha}$ is a \textit{vector} containing the free parameters to be
investigated and $\kappa$ is a normalization constant. This $L$
multiplied by the prior distributions of the parameters (e.g.
eccentricity is restricted to values between 0 to 1) is proportional to
the \textit{a posteriori} probability distribution
$P\left[\hat{\alpha}\right]$ for the model parameters $\hat{\alpha}$.
Since we use no prior information, L is directly proportional to
$P\left[\hat{\alpha}\right]$. Let us note that the logarithm of $L$
coincides with the classic definition of the weighted least-squares so
the maximum of the likelihood function (preferred model) coincides with
the least-squares minimum. 

The resulting $P\left[\hat{\alpha}\right]$ contains more information
than just the optimal model. For example, one can integrate
$P\left[\hat{\alpha}\right]$ to obtain the 68\% confidence level
intervals around the optimal solution. Because the models can be very
complex (e.g. see subsections on the radial velocity and the astrometric
models), an approximate  $P\left[\hat{\alpha}\right]$ typically needs to
be obtained numerically.  Over a highly dimensional parameter space, we
need to efficiently sample $P\left[\hat{\alpha}\right]$ with a list of
$\hat{\alpha}$ values concentrated near the most likely values. The
approach we use to generate such a list is called Monte Carlo Markov Chain
method \citep[MCMC][]{ford:2005} and is a random walk in the parameter
space where the probability of jumping to a new position depends on the
ratio of the new $L$ compared to the current one. To evaluate the
probability of an update, we use the so-called Metropolis-Hastings
approach: the probability of jumping to a new set $\hat{\alpha}_{i+1}$
is 
\begin{equation}
  P_t = \left\{
  \begin{array}{l l}
   1  & \quad {\rm if\,\,} L[\hat{\alpha}_{(i+1)}] > L[\hat{\alpha}_{(i)}]\\
   L[\hat{\alpha}_{(i+1)}]/L[\hat{\alpha}_{(i)} & \quad \rm{otherwise}\\
  \end{array} \right.
\end{equation}
\noindent
and the update is accepted if a random number generated uniformly between $[0,1]$
is smaller than $P_t$. Because this Monte Carlo technique only depends on
the ratio of $L$s, the normalization constant $\kappa$ in
\ref{eq:likelihood} can be ignored in all that follows. The proposed
value of the k-th parameter is obtained by adding a shift $\delta_k$ to
its current value. This $\delta_k$ follow a Gaussian distribution with 0
mean and $\sigma=J_k$, where each $J_k$ should be of the order of the
expected uncertainty on the k-th parameter. If these $J_k$ are too small,
the MCMC will requires many jumps to sample the region of interest. If
$J_k$ is too large, the MCMC will rarely accept updates and the parameter
space will be poorly sampled again. To optimize convergence,
\citet{ford:2005} and other authors found that these $J_k$ need to be
tunned so only 15\% to 30\% of the proposed jumps are accepted. To
perform this optimization, we initialize each $J_k$ using the formal
uncertainty of each parameter obtained from the least-squares solver and
run 10$^5$ MCMC steps. If one parameter has an acceptance rate higher
than 30\%, we multiply $J_k$ by $1.5$. If the acceptance ratio is lower
than 15\% we divide the corresponding $J_k$ by $1.5$. This process is
interated untill all the parameters have acceptance rates between 15\% and
30\%. The precise values of these $J_k$ are not critical as long as the
MCMC converges to the equilibrium distribution. This can be tested by
running different chains and checking that the obtained
$P\left[\hat{\alpha}\right]$ are compatible with each other. Given a
multidimensional parameter space, one needs many steps to properly sample
P$\left[\hat{\alpha}\right]$. This method is computationally expensive
and takes a long time to converge if the Markov Chain is not initialized
close to the favoured solution. As a general rule, we initialize the
Markov Chains within 3-standard deviations as obtained from of the a
least-squares solver on the optimal solution. Once a MCMC is ready and
tuned, we typically run it over 5 $10^6$ steps rejecting the first 10$^5$
to avoid oversampling the initial least-squares solution. The resulting
list of parameters is saved in a file for further processing (e.g.
marginalization, confidence level estimates).

\subsection{Radial velocities only}\label{sec:bayes:rv}

On what follows, all equations used to describe the Keplerian motion and
its observables are based on the expressions given in
\citet{wright:2009}. When only radial velocities are used, the $\chi^2$
function required by equation \ref{eq:likelihood} reads
\begin{eqnarray}
\chi^2_{v}[\hat{\alpha}]=
\sum_{k=1}^{N_{v}} 
\left(\frac{RV_{k}-RV_{\rm model}[t_k;\hat{\alpha}]}{\sigma_k}\right)^2
\end{eqnarray}
\noindent where $RV_{k}$ are the heliocentric RVs, $t_k$ are heliocentric 
instants of observation and $\sigma_k$ are the associated uncertainties. The
predicted radial velocities by the model $RV_{\rm model}$ are given by
\begin{eqnarray}
RV_{\rm model}[t;\alpha]&=& \gamma + K \cos (\omega + \nu(t)) + e \cos \omega \label{eq:rv}
\end{eqnarray}
\noindent where $\nu(t)$ is the so-called true anomaly and is obtained as a
function of time using the relations
\begin{eqnarray}
\tan \frac{\nu(t)}{2} &=& \sqrt{\frac{1+e}{1-e}} \tan \frac{E(t)}{2}\,, \\
E(t) - e \sin E(t) &=& \frac{2\pi}{P} - M_0 \label{eq:kepler}\,.
\end{eqnarray}
\noindent E is the so-called eccentric anomaly and is obtained by solving
numerically the implicit Kepler equation \ref{eq:kepler}
\footnote{\texttt{http://mathworld.wolfram.com/KeplersEquation.html}}.
The constant $K$ is the radial velocity semiamplitude and relates to the
physical parameters of the system as follows
\begin{eqnarray}
K^3 =  \frac{2\pi G}{P \left(1-e^2\right)^{3/2}} 
\, \frac{m^3 \sin^3 i}{\left(M_*+m_p\right)^2}
\end{eqnarray}
\noindent where G is the gravitational constant in MKS units. The
physical free parameters $\hat{\alpha}$ in the condition equation
\ref{eq:rv} to be solved are
\begin{itemize}
\item $\gamma$ : Systemtic radial velocity of the star 
(or radial velocity offset) in m s$^{-1}$
\item m $\sin i$ : Planet mass times the orbital inclination (or minimum mass) in Kg
\item P : orbital period in seconds
\item e : orbital eccentricity (from 0 to 1 for bound orbits).
\item $\omega$ : argument of the periastron in radians
\item $M_0$ : initial mean anomaly in radians.
\end{itemize}
\noindent When only dealing with radial velocity measurements,the
least-squares solvers and the Bayesian MCMC directly optimize
on the parameters listed above.

\subsection{Astrometry and radial velocities}\label{sec:bayes:astro}

When two-dimensional astrometric measurements $u,v$ and radial velocities are
available, the $\chi^2$ function required by equation \ref{eq:likelihood} reads
\begin{eqnarray}
\chi^2 &=&
\sum_k^{N_{\rm RV}} 
\left(\frac{RV_{\rm obs}-RV_{\rm model}[t_k\hat{\alpha}]}{\sigma_k}\right)^2 \\
&+&
\sum_s^{N_{\rm astro}} 
\left(\frac{u_{\rm s}-u_{\rm model}[t_s;\hat{\alpha}]}{\sigma_s}\right)^2 +
\sum_s^{N_{\rm astro}} 
\left(\frac{v_{\rm s}-v_{\rm model}[t_s;\hat{\alpha}]}{\sigma_s}\right)^2
\end{eqnarray}
\noindent where $RV_{\rm model}[t_k\hat{\alpha}]$ is given in \ref{eq:rv}. The
astrometric condition equations in the local tangent plane coordinates read
\begin{eqnarray}
u[t;\hat{\alpha}] &=& u_0 + \mu_{\alpha} \left(t-t_0\right) - \Pi p_{\alpha}(t) + \Delta_\alpha(t)\, \\
v[t;\hat{\alpha}] &=& v_0 + \mu_{\delta} \left(t-t_0\right) - \Pi p_{\delta}(t)+ \Delta_\delta(t)\, ,
\end{eqnarray}
\noindent where all the linear parameters $u_0,v_0,\mu_\alpha,\mu_\delta$ and
$\Pi$  are already described in \ref{sec:processing}. The Keplerian parameters
are included in $\Delta_\alpha$ and $\Delta_\alpha$ as
\begin{eqnarray}
\Delta_\alpha(t) &=& B X(t) + G Y(t) \label{eq:astrox}\\
\Delta_\delta(t) &=& A X(t) + F Y(t) \label{eq:astroy}\\
\nonumber \\
A &=& a (\cos \Omega \cos \omega - \sin \Omega \sin \omega \cos i) \\
B &=& a (\sin \Omega \cos \omega + \cos \Omega \sin \omega \cos i) \\
F &=& a (-\cos \Omega \sin \omega - \sin \Omega \cos \omega \cos i) \\
G &=& a (-\sin \Omega \sin \omega + \cos \Omega \cos \omega \cos i)
\end{eqnarray}
\noindent where A, B, F and G are the so-called Thiele Innes constants.
$a$ is the orbital semi-major axis and is related to the other
parameters as
\begin{eqnarray}
\left(\frac{a}{AU}\right)^3 = \frac{(\Pi/1000)^3 (m/M_{\sun})^3}{((M_*+m)/M_{\sun})^2}\, (P/yr)^2\,, 
\end{eqnarray}
\noindent where $M_\sun$ is the mass of the sun in Kg and $yr$ is a year in
seconds. All the time dependence in the astrometric motion in \ref{eq:astrox} and
\ref{eq:astroy} is in $X(t)$ and $Y(t)$. X and Y represent the Keplerian motion
of the star on a coordinate system coplanar with the orbital plane with the X
axis pointing to the orbital periastron. X and Y depend on time through the
eccentric anomaly E as
\begin{eqnarray}
X(t) &=& \cos E(t) - e\\
Y(t) &=& \sqrt{1-e^2} \sin E(t)
\end{eqnarray}
\noindent and E has to be solved as for the radial velocities using
\ref{eq:kepler}. The free parameters $\hat{\alpha}$ for the combined
astrometric and radial velocity measurements are
\begin{itemize}
\item $\gamma$ : Systemtic radial velocity of the star, 
(or radial velocity offset) in m s$^{-1}$
\item m : Planet mass in Kg.
\item P : Orbital period in seconds
\item e : Orbital eccentricity (from 0 to 1 for bound orbits)
\item i : Orbital inclination with respect the plane of the sky in radian (0 corresponds to
a face-on orbit).
\item $\Omega$ : Argument of the node in radians (orientation of the orbit on the
sky with respect to the local North. Positive from North to East).
\item $\omega$ : Argument of the periastron in radians. This is the angle between
the node and the periastron of the system as measured on the orbital plane.
\item $M_0$ : initial mean anomaly in radians.
\item $u_0$ : offset in R.A. in mas.
\item $v_0$ : offset in Dec. in mas.
\item $\mu_\alpha$ : proper motion in R.A. in mas yr$^{-1}$.
\item $\mu_\delta$ : proper motion in Dec. in mas yr$^{-1}$.
\item $\Pi $ : parallax in mas.
\end{itemize}
\noindent A more detailed description of these parameters can be found elsewhere.
This prescription is based on the definitions given by \citep{wright:2009}.

\begin{deluxetable}{lcccccccccc}
\tabletypesize{\scriptsize}
\rotate
\tablecaption{Reference frame stars}\label{tab:references}
\tablewidth{0pt}
\tablehead{
 \colhead{CAPS ID} & 
 \colhead{ NOMAD ID   } &  
 \colhead{ R.A.}& 
 \colhead{ Dec }& 
 \colhead{ $\Pi_{obs}$ }& 
 \colhead{ $\epsilon_{\Pi}$} & 
 \colhead{ $\mu_{\alpha}$ }& 
 \colhead{ $\epsilon_{\mu\alpha}$ }& 
 \colhead{ $\mu_{\delta}$} & 
 \colhead{ $\epsilon_{\mu\delta}$ }&
 \colhead{ RMS}  \\
\colhead{		   } &
\colhead{		   } &
\colhead{ (deg) 	   } &
\colhead{ (deg) 	   } &
\colhead{ (mas) 	   } &
\colhead{ (mas) 	   } &
\colhead{ (mas yr$^{-1}$)  } &
\colhead{ (mas yr$^{-1}$)  } &
\colhead{ (mas yr$^{-1}$)  } &
\colhead{ (mas yr$^{-1}$)  } &
\colhead{ (mas) 	     } 
}
\startdata
29 &  0665-0207029 & 130.2219847 & -23.4573308 &  0.346  &  0.450  & -1.377  &   0.213  &    4.566  &	0.409 &   0.522  \\
37 &  0665-0207087 & 130.2419722 & -23.4835250 &  0.190  &  0.430  &  5.762  &   0.207  &   -3.339  &	0.794 &   0.721  \\
42 &  0665-0207030 & 130.2221722 & -23.4462111 & -0.241  &  0.470  & -1.260  &   0.525  &    2.049  &	1.043 &   0.825  \\  
47 &  0665-0207068 & 130.2353750 & -23.4446139 &  0.591  &  0.485  &  3.457  &   0.321  &   -4.699  &	0.662 &   0.727  \\  
49 &  0665-0207065 & 130.2348028 & -23.4468139 &  0.089  &  0.289  &  4.129  &   0.357  &   -1.206  &	0.404 &   0.860  \\  
51 &  0665-0207007 & 130.2134861 & -23.4580083 &  1.998  &  0.308  & -8.148  &   0.270  &  -10.953  &	0.307 &   0.778  \\
52 &  0665-0207077 & 130.2391917 & -23.4666139 & -0.017  &  0.290  & -1.409  &   0.221  &    0.267  &	0.319 &   0.800  \\  
54 &  0665-0207140 & 130.2531583 & -23.4349056 &  0.290  &  0.490  & -0.429  &   0.531  &    2.920  &	1.601 &   0.976  \\  
74 &  0665-0207043 & 130.2277000 & -23.4282472 &  0.017  &  0.316  & -0.756  &   0.279  &   -1.585  &	0.368 &   0.630  \\
76 &  0665-0207200 & 130.2732667 & -23.4496417 &  1.318  &  0.830  &  0.694  &   3.097  &    0.477  &   7.131 &   0.603  \\
78 &  0665-0207195 & 130.2712806 & -23.4570806 &  0.083  &  0.599  &  4.436  &   0.317  &   -2.147  &	0.932 &   0.542  \\ 
\enddata
\end{deluxetable}

\begin{deluxetable}{lccccccccccccc}
\tabletypesize{\scriptsize}
\rotate
\tablecaption{Stars used to estimate the parallax zero-point. 
The zero-point corrections is 
the average $\Pi_{obs} -\Pi_{phot}$ and ammounts to
$0.23 \pm 0.1$.
}\label{tab:zeropoint}
\tablewidth{0pt}
\tablehead{
\colhead{  CAPS ID    }&  
\colhead{  $ \mu_{\alpha} $ }& 
\colhead{  $ \epsilon_{\mu\alpha}$ }& 
\colhead{  $ \mu_{\delta} $ }& 
\colhead{  $ \epsilon_{\mu\delta}$ }&
\colhead{  B }& 
\colhead{  J }& 
\colhead{  H }& 
\colhead{  K }& 
\colhead{  Phot. dist }&
\colhead{  $\Pi_{phot}$ }&
\colhead{  $\Pi_{phot} -\Pi_{obs}$ }&
\colhead{  T$_{eff}$} &
\colhead{  Sp.Type}\\
  & 
\colhead{ (mas yr$^{-1}$) }& 
\colhead{ (mas yr$^{-1}$) }& 
\colhead{ (mas yr$^{-1}$) }& 
\colhead{ (mas yr$^{-1}$) }& 
\colhead{ (mag) }& 
\colhead{ (mag) }& 
\colhead{ (mag) }& 
\colhead{ (mag) }& 
\colhead{ (pc) }&
\colhead{ (mas) }&
\colhead{ (mas) }&
\colhead{ (K) }& 
\colhead{ }  
   }
\startdata
29 & -10.1   &  5.0   &  3.4  &    4.8  & 15.750  &  14.554 & 14.228 & 14.251 & 2272 & 0.4400 &  0.0940 &  6496  &   F5V  \\
42 &  0.0   &  0.0   &  0.0  &    0.0  & 16.080  &  15.394 & 15.071 & 14.871  & 4900 & 0.2041 &  0.4451 &  6717  &   F2V  \\
47 &  0.0   &  0.0   &  0.0  &    0.0  & 17.140  &  15.327 & 14.897 & 14.468  & 1527 & 0.6548 &  0.0638 &  5347  &   G9V  \\
49 &  0.0   &  0.0   &  0.0  &    0.0  & 17.220  &  15.289 & 14.660 & 14.900  & 1906 & 0.5244 &  0.4354 &  5635  &   G6V  \\
52 &-12.0   &  5.0   & 26.0  &    10.0 & 16.120  &  15.455 & 15.105 & 14.871  & 4781 & 0.2091 &  0.2261 &  6687  &   F3V  \\
54 &  0.0   &  0.0   &  0.0  &    0.0  & 17.200  &  15.698 & 15.503 & 15.337  & 3739 & 0.2674 & -0.0226 &  6130  &   F8V  \\
78 &  0.0   &  0.0   &  0.0  &    0.0  & 18.300  &  15.980 & 15.952 & 15.628  & 2063 & 0.4847 &  0.4017 &  5347  &   G9V  \\
\enddata
\end{deluxetable}

\end{document}